\documentclass[12pt]{article}
\usepackage{braket}
\usepackage{scicite}
\usepackage{graphicx}
\usepackage{times}
\usepackage{amsmath}
\usepackage{amssymb}
\usepackage{floatrow}
\newenvironment{sciabstract}{%
\begin{quote} \bf}
{\end{quote}}
\title{Unraveling electronic correlations in warm dense quantum plasmas} 
\author
{T.~Dornheim,$^{1,2\ast}$ T.~D\"oppner,$^{3}$ P.~Tolias,$^{4}$  \\ M.~P.~B\"ohme,$^{1,2,5}$ L.B.~Fletcher,$^6$  Th. Gawne,$^{1,2}$  F.~R.~Graziani,$^{3}$\\ D.~Kraus,$^{7,2}$  M.~J.~MacDonald,$^3$ Zh.~A.~Moldabekov,$^{1,2}$\\  S. Schwalbe,$^{1,2}$  D.O.~Gericke,$^8$ and J.~Vorberger$^{2}$  \\  
\\
\normalsize{$^{1}$Center for Advanced Systems Understanding (CASUS), D-02826 G\"orlitz, Germany}\\
\normalsize{$^{2}$Helmholtz-Zentrum Dresden-Rossendorf (HZDR), D-01328 Dresden, Germany}\\
\normalsize{$^{3}$Lawrence Livermore National Laboratory (LLNL), California 94550 Livermore, USA}\\
\normalsize{$^{4}$Space and Plasma Physics, Royal Institute of Technology (KTH)}\\
\normalsize{Stockholm, SE-100 44, Sweden}\\
\normalsize{$^{5}$Technische  Universit\"at  Dresden,  D-01062  Dresden,  Germany}\\
\normalsize{$^{6}$SLAC National Accelerator Laboratory, CA 94025 Menlo Park, USA}\\
\normalsize{$^{7}$Institut f\"ur Physik, Universit\"at Rostock, D-18057 Rostock, Germany}\\
\normalsize{$^{8}$Centre for Fusion, Space and Astrophysics, University of Warwick, Coventry CV4 7AL, UK}\\
\\
\normalsize{$^\ast$To whom correspondence should be addressed; E-mail:  t.dornheim@hzdr.de}
}
\date{}
\begin{document} 
\baselineskip24pt
\maketitle 

\begin{sciabstract}
The study of matter at extreme densities and temperatures 
has emerged as a highly active frontier at the interface of plasma physics, material science and quantum chemistry with direct relevance for planetary modeling and inertial confinement fusion. 
 A particular feature of such warm dense matter is the complex interplay of strong Coulomb interactions, quantum effects, and thermal excitations, rendering its rigorous theoretical description a formidable challenge.
 Here, we report a breakthrough in path integral Monte Carlo simulations that allows us to unravel this intricate interplay for light elements without nodal restrictions. 
 This new capability gives us access to electronic correlations previously unattainable. 
 As an example, we apply our method to
 strongly compressed beryllium to describe
 x-ray Thomson scattering (XRTS) data obtained at the National Ignition Facility. 
We find excellent agreement between simulation and experiment. 
Our analysis shows an unprecedented level of consistency for independent observations without the need for any empirical input parameters.
\end{sciabstract}

Matter at extreme densities and temperatures displays a complex quantum behavior.
A particularly intriguing situation emerges when the interaction, thermal, and Fermi energies are comparable.
Understanding such warm dense matter (WDM) 
requires a holistic description taking into account partial ionization, partial quantum degeneracy, and moderate coupling. Indeed, even familiar concepts such as well-defined electronic bound states and ionization break down in this regime.

Interestingly, such conditions are widespread throughout the universe, naturally occurring in a host of astrophysical objects such as giant planet interiors~\cite{Kraus_Science_2022}, brown and white dwarfs~\cite{Kritcher_Nature_2020}, and, on Earth, meteor impacts~\cite{Hanneman_Science_1967}.
Moreover, WDM plays a key role in cutting-edge technological applications such as 
the discovery and synthesis of novel materials~\cite{Kraus2016}.
An extraordinary achievement has recently been accomplished in the field of inertial confinement fusion at the National Ignition Facility (NIF)~\cite{Betti2016,Zylstra2022}. 
In these experiments, both the ablator and the fuel traverse the WDM regime, making a rigorous understanding of such states paramount to reach the reported burning plasma and net energy gain~\cite{Lawson_PRL_2022,Ignition_PRL_2024}.

The pressing need to understand extreme states has driven a large leap in the experimental capabilities; the considerable number of remarkable successes includes the demonstration of diamond formation under planetary conditions~\cite{Kraus2016,Kraus2017}, opacity measurements under solar conditions~\cite{Bailey2015}, probing atomic physics at Gigabar pressures~\cite{Tilo_Nature_2023}, and the determination of energy loss of charged particles~\cite{Malko_NatComm_2022,Cayzac2017}.
However, this progress is severely hampered: 
to diagnose WDM experiments, a thorough understanding of the electronic response is indispensable. Indeed, even the inference of basic parameters such as temperature and density requires rigorous modeling to interpret the probe signal.

Density functional theory combined with classical molecular dynamics for the ions (DFT-MD) has emerged as the de-facto work horse for computing WDM properties. 
While being formally exact~\cite{Mermin_DFT_1965}, the predictive capability of DFT-MD is limited by the unknown exchange--correlation functional, which has to be approximated in practice, and the application of the Born-Oppenheimer approximation. 
A potentially superior alternative is given by \emph{ab initio} path integral Monte Carlo (PIMC) simulations~\cite{cep}, which are in principle capable of providing an exact solution for a given quantum many-body problem without any empirical input. Yet, PIMC simulations of quantum degenerate Fermi systems, such as the electrons in WDM, are afflicted with an exponential computational bottleneck, which is known as the fermion sign problem~\cite{troyer,dornheim_sign_problem}. As a consequence, PIMC application to matter under extreme conditions has either been limited to comparably simple systems such as the uniform electron gas model~\cite{review}
, or based on inherently uncontrolled approximations as in the case of restricted PIMC~\cite{Brown_PRL_2013,Militzer_PRL_2015,Militzer2021EOS}.

Here, we present a solution to this unsatisfactory situation and demonstrate its capabilities on the example of warm dense beryllium (Be).
 Since our approach is not based on any nodal restriction, we get access to the full spectral information in the imaginary-time domain~\cite{Dornheim_review}. As the capstone of our work, we employ our simulations to re-analyze 
 X-ray Thomson scattering (XRTS) data obtained at the NIF for strongly compressed Be in a backscattering geometry~\cite{Tilo_Nature_2023}. In addition, we consider a new data set that has been measured at a smaller scattering angle that is more sensitive to electronic correlations. 
 Our unique access to electron correlation functions allows for novel ways to interpret the XRTS data, resulting in an unprecedented level of consistency. 
We are convinced that our work will open up a wealth of new avenues for substantial advances of our understanding of warm dense quantum plasmas.

\begin{figure*}\centering\includegraphics[width=1\textwidth]{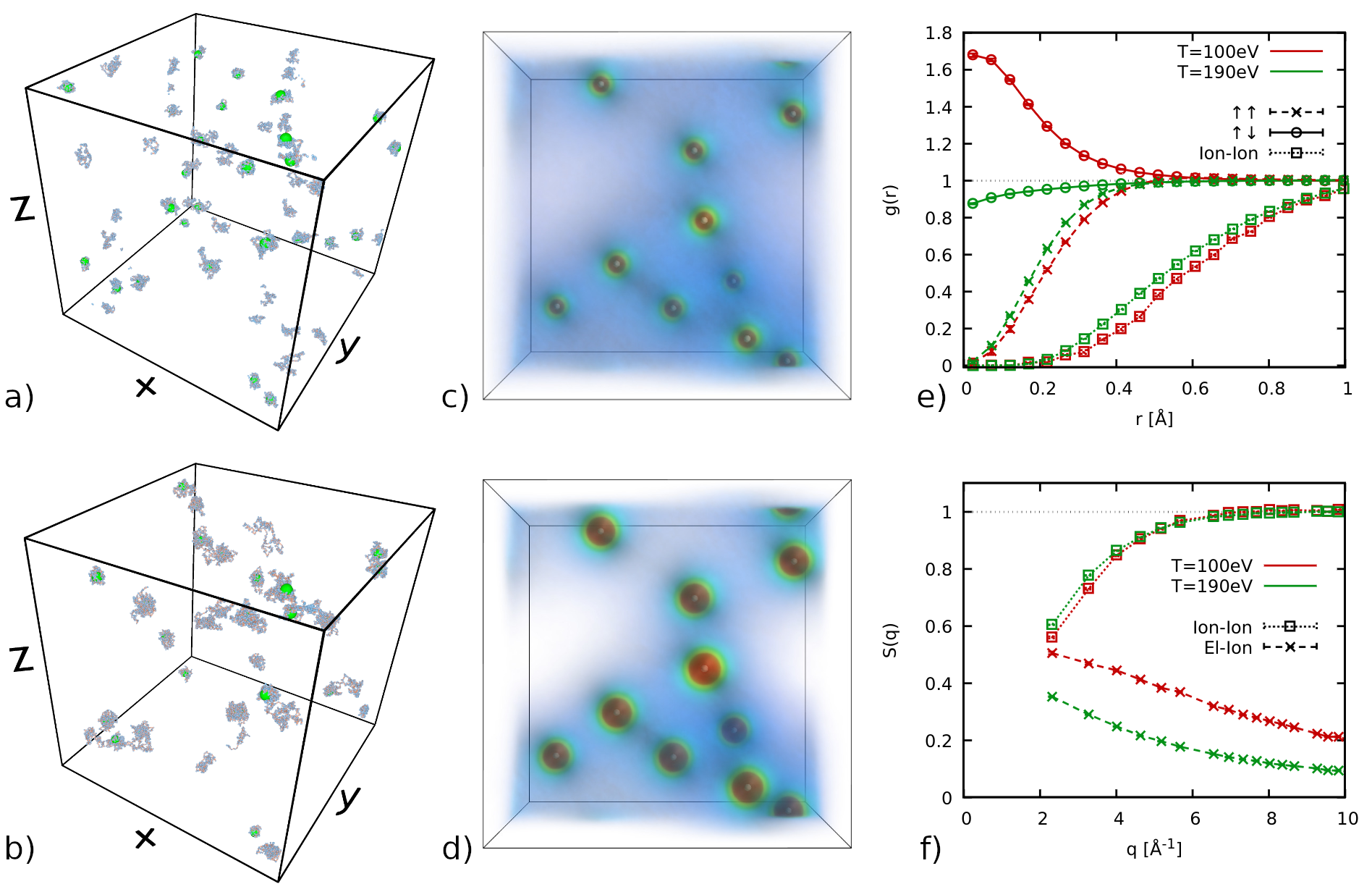}
\caption{\label{fig:PIMC}
\textit{Ab initio} PIMC simulations of compressed Be ($\rho=7.5\,$g/cc). a) Snapshot of a PIMC simulation of $N_\textnormal{Be}=25$ Be atoms at $T=190\,$eV. b) same as a) but for  $T=100\,$eV. The green orbs show the ions and the blue-red paths the quantum degenerate electrons; c),d) electronic density in real space for a fixed ion configuration at the same temperatures. e) Our PIMC simulations give us access to all many-body correlations in the systems, including the spin-resolved electron--electron pair correlation functions $g_{\uparrow\uparrow}(r)$ and $g_{\uparrow\downarrow}(r)$, and the ion--ion pair correlation function $g_{II}(r)$. f) Electron--ion and ion--ion static structure factors $S_{eI}(q)$ and $S_{II}(q)$, giving us access to the ratio of elastic and inelastic contributions to the full scattering intensity, see the main text.
}
\end{figure*} 

\paragraph*{Simulation approach.} In principle, the PIMC method allows one to obtain an exact solution to the full quantum-many body problem without any empirical input or approximations. However, the application of PIMC to quantum degenerate electrons is severely hampered by the fermion sign problem~\cite{troyer,dornheim_sign_problem}. To circumvent this obstacle, Militzer, Ceperley and others~\cite{Brown_PRL_2013,Militzer_PRL_2015,Militzer2021EOS,Ceperley1991}
have employed the \emph{fixed-node approximation}. This restricted PIMC method allows for simulations of large systems without a sign problem, an advantage that comes at the cost of a de-facto uncontrolled approximation~\cite{Schoof_PRL_2015,Malone_PRL_2016}. Moreover, the implementation of a nodal restriction prevents the usual access of PIMC to the full spectral information in the imaginary-time domain~\cite{Dornheim_insight_2022,Dornheim_review}, preventing a direct comparison with XRTS measurements.

In this work, we employ a fundamentally different strategy by carrying out a controlled extrapolation over a continuous variable $\xi\in[-1,1]$ that is substituted into the canonical partition function~\cite{Xiong_JCP_2022,dornheim2023fermionic,Dornheim_JPCL_2024}, see the Supplemental Material~\cite{supplement}. This treatment removes the exponential scaling of the computation time with the system size for substantial parts of the WDM regime without the need for any empirical input such as the nodal surface of the density matrix for restricted PIMC or the XC functional for DFT. At the same time, it retains full access to the spectral information about the system encoded in the imaginary-time density-density correlation function (ITCF), thereby allowing for direct comparison between simulations and XRTS measurements. While this approach had been successfully applied to the uniform electron gas model~\cite{dornheim2023fermionic,Dornheim_JPCL_2024}, we use it here to study the substantially more complex case of electrons and nuclei in WDM 
for the first time.

In Figs.~\ref{fig:PIMC}a) and b), we show snapshots of all-electron PIMC simulations of $N_\textnormal{Be}=25$ Be atoms (i.e, $N_\textnormal{e}=100$ electrons) for the mass density $\rho=7.5\,$g/cc and temperatures of $T=190\,$eV and $T=100\,$eV, respectively. The green orbs depict the nuclei, which behave basically as classical point particles, although this is not built into our simulations. The blue paths represent the quantum degenerate electrons; their extension is proportional to the thermal de Broglie wavelength $\lambda_T=\hbar\sqrt{2\pi/m_\textnormal{e}k_\textnormal{B}T}$
and serves as a measure for the average extension of a hypothetical single-electron wave function. The interplay of electron delocalization with effects such as Coulomb coupling shapes the physical behavior of the system.
In panels c) and d), we illustrate PIMC results for the spatially resolved electron density in the external potential of a fixed ion configuration. We find a substantially increased localization around the nuclei for the lower temperature.

Figs.~\ref{fig:PIMC}e) and f) show \emph{ab initio} PIMC results for the full Be system, where both electrons and nuclei are treated dynamically on the same level. 
Specifically, panel e) shows various pair correlation functions, where the red and green lines correspond to $T=100\,$eV and $T=190\,$eV, respectively.
The ion--ion pair correlation function $g_{II}(r)$ [squares] is relatively featureless in both cases. The same holds for the spin-diagonal electron--electron pair correlation function $g_{\uparrow\uparrow}(r)=g_{\downarrow\downarrow}(r)$ [crosses], although the exchange--correlation hole is substantially reduced when compared to $g_{II}(r)$ mainly due to the weaker Coulomb repulsion. In stark contrast, the spin-offdiagonal pair correlation function $g_{\uparrow\downarrow}(r)$ [circles] exhibits a nontrivial behavior and strongly depends on the temperature. While being nearly flat for $T=190\,$eV, $g_{\uparrow\downarrow}(r)$ markedly increases towards $r=0$ for $T=100\,$eV. This increased contact probability is a direct consequence of the substantial presence of ions with a fully occupied K-shell at the lower temperature, and nicely illustrates the capability of our PIMC simulations to capture the complex interplay of ionization, thermal excitation, and electron--electron correlations.
Finally, panel f) shows corresponding results for the ion--ion [squares] and electron-ion [crosses] static structure factor (SSF). These contain important information about the generalized form factor and Rayleigh weight, which are key properties in the interpretation of XRTS experiments~\cite{Tilo_Nature_2023} and a gamut of other applications. 

\begin{figure*}\centering\vspace*{-0cm}
\includegraphics[width=1\textwidth]{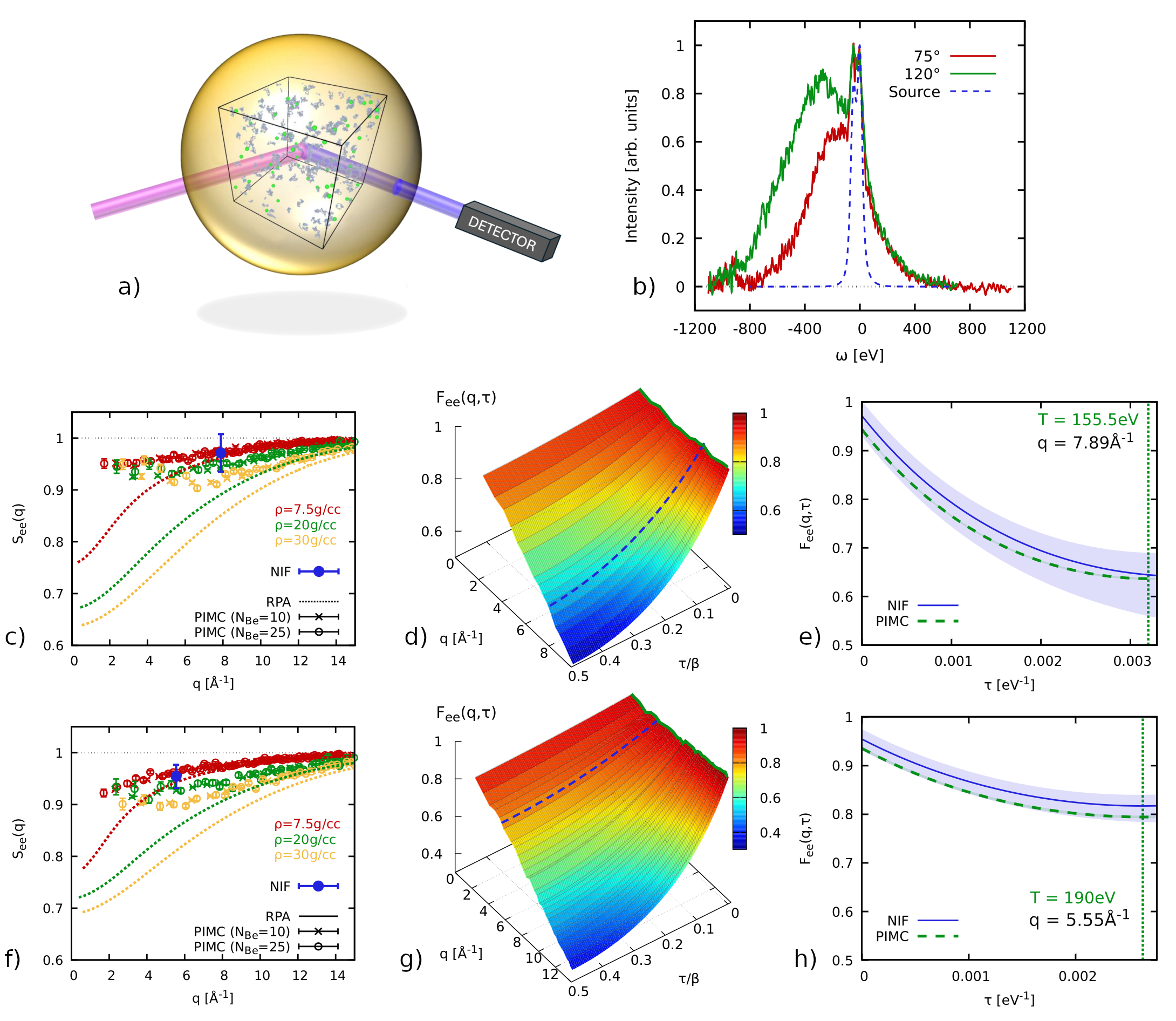}
\caption{\label{fig:results} a) Schematic illustration of our setup. The Be capsule is compressed and probed by an x-ray source (purple); the scattered photons (blue) are collected by a detector under an angle $\theta$ with respect to the incident beam. We use the PIMC method to simulate the quantum degenerate interior of the capsule, allowing for unprecedented comparisons between theory and experiment; b) XRTS spectra for $\theta=120^\circ$ (green) and $\theta=75^\circ$ (red), and the source-and-instrument function (blue dashed); c,f) PIMC results for $S_{ee}(q)$ for different densities (symbols) compared to the NIF data point (bold blue) and random phase approximation (RPA) results (dotted lines); d,g) ITCF $F_{ee}(q,\tau)$ in the $q$-$\tau$-plane, with the colored surface and dashed blue line corresponding to PIMC simulations for $\rho=20\,$g/cc and the Laplace transform of the NIF spectra; e,h) $\tau$-dependence of $F_{ee}(q,\tau)$ at the probed wavenumber. The center and bottom rows correspond to the $\theta=120^\circ$ and $\theta=75^\circ$ shots, for which we find $T=155.5\,$eV and $T=190\,$eV (see the vertical dotted lines in e,h), respectively~\cite{supplement}.
}
\end{figure*} 

\paragraph*{Results.} As a demonstration of our new PIMC capabilities, we re-analyze an XRTS experiment with strongly compressed beryllium at the National Ignition Facility~\cite{Tilo_Nature_2023} and repeated the experiment at a smaller scattering angle to focus more explicitly on electronic correlation effects.
Fig.~\ref{fig:results}a) shows an illustration of the experimental set-up using the GBar XRTS platform. 
$184$ optical laser beams (not shown) are used for the hohlraum compression~\cite{Tilo_Nature_2023} of a Be capsule (yellow sphere) which is filled with a core of air. A further $8$ laser beams are used to heat a zinc foil generating $8.9\,$keV X-rays~\cite{MacDonald_POP_2022} that are used to probe the system (purple ray).
By detecting the scattered intensity (blue ray) at an angle $\theta$, we get insight into the microscopic physics of the sample on a specific length scale; the same microscopic physics can be resolved by our new PIMC simulations, a snapshot of which is depicted inside the Be capsule.

The measured XRTS spectra are shown as the green and red curves in Fig.~\ref{fig:results}b) and have been obtained at scattering angles of $\theta=120^\circ$~\cite{Tilo_Nature_2023} and $\theta=75^\circ$ (new). 
They are given by a convolution of the dynamic structure factor $S_{ee}(q,\omega)$ with the combined source-and-instrument function $R(\omega)$ [dashed blue]. Since a deconvolution to extract $S_{ee}(q,\omega)$ is unstable,
we instead perform a two-sided Laplace transform~\cite{Dornheim_T_2022,Dornheim_T2_2022,Dornheim_review}
\begin{eqnarray}\label{eq:laplace_main_text}
    F_{ee}(q,\tau) = \mathcal{L}\left[S_{ee}(q,\omega)\right] = \int_{-\infty}^\infty \textnormal{d}\omega\ S_{ee}(q,\omega) e^{-\hbar\omega\tau}\ ;
\end{eqnarray}
the well-known convolution theorem then gives us direct access to the imaginary-time correlation function (ITCF) $F_{ee}(q,\tau)$ based on the experimental data, with $\tau\in[0,\beta]$ being the imaginary time and $\beta=1/k_\textnormal{B}T$ the inverse temperature~\cite{supplement}. The ITCF contains the same information as $S_{ee}(q,\omega)$, but in a different representation~\cite{Dornheim_insight_2022}.
A particularly important application of $F_{ee}(q,\tau)$ is the model-free estimation of the temperature~\cite{Dornheim_T_2022,Dornheim_T2_2022}, and we find $T=155.5\,$eV and $T=190\,$eV for $\theta=120^\circ$ and $\theta=75^\circ$, respectively.

A second advantage of Eq.~(\ref{eq:laplace_main_text}) is that it facilitates the direct comparison of the experimental observation with our new PIMC results. 
As a first point, we consider the electronic static structure factor $S_{ee}(q)=F_{ee}(q,0)$ in Fig.~\ref{fig:results}c) and f) for the two relevant temperatures, and the circles and crosses show PIMC results for $N_\textnormal{Be}=25$ and $N_\textnormal{Be}=10$ beryllium atoms.
Evidently, no finite-size effects can be resolved within the given error bars with the possible exception of the smallest $q$ values.
A particular strength of the PIMC method is that it allows us to unambiguously resolve the impact of electronic XC-effects. To highlight their importance for the description of warm dense quantum plasmas even in the high-density regime, we compare the PIMC data with the mean-field based random phase approximation~\cite{kraeft2012quantum} (dotted lines). 
The latter approach systematically underestimates the true $S_{ee}(q)$ and only becomes exact in the single-particle limit of large wave numbers.
The blue circles correspond to $S_{ee}(q)$ extracted from the NIF data following the procedure introduced in the recent Ref.~\cite{dornheim2023xray}. They are consistent with the PIMC results for $\rho\lesssim20\,$g/cc.

The full ITCF $F_{ee}(q,\tau)$ is shown in panels d) and g) in the $q$-$\tau$-plane, where the coloured surface shows the PIMC results for $\rho=20\,$g/cc, and the dashed blue lines have been obtained from the experimental data via a two-sided Laplace transform, see Supplemental Material~\cite{supplement}.
Clearly, the ITCF exhibits a rich structure that is mainly characterized by an increasing decay with $\tau$ for larger wave numbers. In fact, this $\tau$-dependence is a direct consequence of the quantum delocalization of the electrons~\cite{Dornheim_insight_2022,Dornheim_PTR_2022} and would be absent in a classical system. 
The NIF data are in excellent agreement with our PIMC simulations over the entire $\tau$-range.
This can be seen particularly well in panels e) and h), where we show the ITCF for the fixed values of $q$ probed in the experiment.
We find a more pronounced decay of $F_{ee}(q,\tau)$ with increasing $\tau$ for larger $q$. 
A second effect is driven by the different temperatures of these separate NIF shots, as a higher temperature leads to a reduction of quantum delocalization and, therefore, a reduced $\tau$-decay.
The observed agreement between the PIMC results and the experimental data for different $q$ and temperature is thus nontrivial and 
constitutes a remarkable level of agreement and consistency between theory and experiment.

\begin{figure}\centering
\includegraphics[width=1\textwidth]{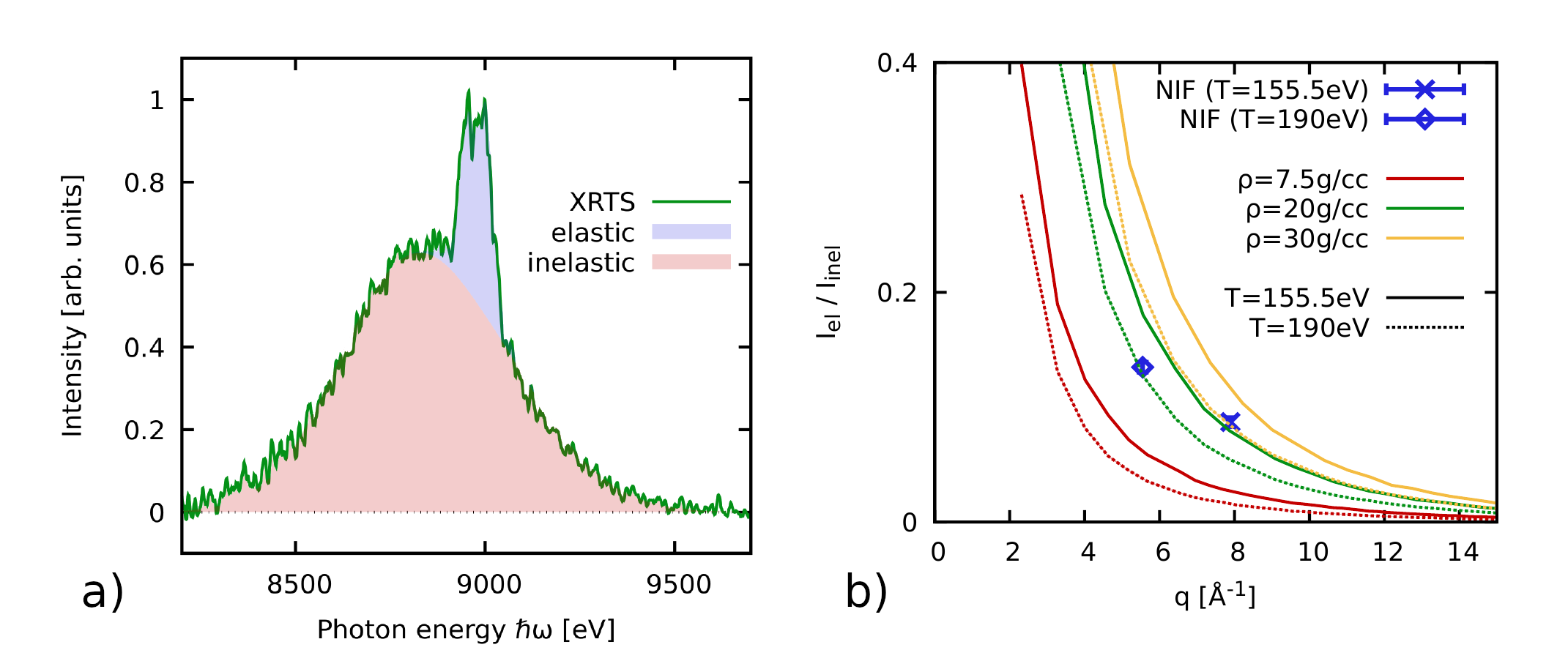}
\caption{\label{fig:form_factor}
a) Extracting the ratio of elastic (blue) and inelastic (red) contributions $I_\textnormal{el}/I_\textnormal{inel}$ from the XRTS measurement at $\theta=75^\circ$ ($q=5.55\,$\AA$^{-1}$); b) wavenumber dependence of the ratio $I_\textnormal{el}/I_\textnormal{inel}$. Solid (dotted) lines: PIMC results for $T=155.5\,$eV ($T=190\,$eV) for $\rho=7.5\,$g/cc (red), $\rho=20\,$g/cc (green), and $\rho=30\,$g/cc (yellow); blue cross and diamond: NIF measurements for $\theta=120^\circ$ and $\theta=75^\circ$.
}
\end{figure}

Finally, we consider an additional observable that can be directly extracted from the experimental data: the ratio of the elastic to the inelastic contributions to the full scattering intensity $I_\textnormal{el}/I_\textnormal{inel}$.
In Fig.~\ref{fig:form_factor}a), the two components are illustrated for the case of $\theta=75^\circ$. In practice, the elastic signal has the form of the source function [cf.~Fig.~\ref{fig:results}b)], and $I_\textnormal{inel}$ is given by the remainder.
The ratio $I_\textnormal{el}/I_\textnormal{inel}$ constitutes a distinct measure for the localization of the electrons around the ions on the probed length scale determined by $q$. Therefore, it is highly sensitive to system parameters such as the density, and, additionally, to the heuristic but often useful concept of an effective ionization degree~\cite{Tilo_Nature_2023}.
Yet, the prediction of $I_\textnormal{el}/I_\textnormal{inel}$ from \emph{ab initio} simulations requires detailed knowledge about correlations between all particle species, see the Supplemental Material~\cite{supplement}. This requisition is beyond the capabilities of standard DFT-MD, but straightforward with our new PIMC simulations, cf.~Fig.~\ref{fig:PIMC}.

In Fig.~\ref{fig:form_factor}b), we show our simulation results for $I_\textnormal{el}/I_\textnormal{inel}$ for two temperatures and three densities.
The comparison with the experimental data points yields excellent agreement with $\rho=20\,$g/cc for both experiments,
which is fully consistent with the independent analysis of $S_{ee}(q)$ presented in Fig.~\ref{fig:results}.
Moreover, Fig.~\ref{fig:form_factor}b) further substantiates the difference in temperature in the two separate NIF shots that we have observed from the model-free ITCF method~\cite{supplement}.
This nicely illustrates the unprecedented degree of consistency in the analysis of XRTS signals facilitated by our new simulation capabilities.

\paragraph*{Discussion.} We have presented a novel framework for the highly accurate \emph{ab initio} PIMC simulation of warm dense quantum plasmas, treating electrons and ions on the same level. As an application, we have analyzed XRTS measurements of strongly compressed Be using existing data~\cite{Tilo_Nature_2023}
as well as a new data set that probes larger length scales where electronic XC-effects are more important.
Due to their unique access to electronic correlation functions, our PIMC simulations have allowed us to independently analyze various aspects of the XRTS signal such as the ITCF $F_{ee}(q,\tau)$ and the ratio of elastic to inelastic contributions.
We have thus demonstrated a remarkable consistency between simulation and experiment without the need for any empirical parameters.

Our PIMC simulations accurately capture phenomena that manifest over distinctly different length scales due to the simulation of potentially hundreds of electrons and nuclei~\cite{Dornheim_JPCL_2024}.
This is particularly important for upcoming XRTS measurements with smaller scattering angles, and for the description of properties that can be probed in the optical limit such as electric conductivity and reflectivity. A key strength is our capability to resolve any type of many-particle correlation function between either electrons or ions. This is in stark contrast with standard DFT-MD simulations, where the computation of electronic correlation functions is not possible even if the exact XC functional were known. Moreover, our PIMC simulations are not confined to two-particle correlation functions and linear response properties such as the dynamic structure factor probed in XRTS~\cite{Dornheim_review}. 

Our highly accurate PIMC results will spark a host of developments in the simulation of WDM. Most importantly, they can unambiguously benchmark the accuracy of existing DFT approaches,
and provide crucial input for the construction of advanced nonlocal XC functionals~\cite{pribram}. 
In addition, our results can quantify the nodal error in the restricted PIMC approach, support dynamic methods including time-dependent DFT, and test the basic underlying assumptions in widely used theoretical models.

Finally, our simulations will have a direct and profound impact on nuclear fusion and astrophysics.
Due to their dependable predictive capability, they provide both key input for integrated modeling such as transport properties and the equation of state and guide the development of experimental set-ups. 
A case in point is given by XRTS measurements, for which we have demonstrated the capabilities of our PIMC approach to give a highly consistent interpretation of the data for strongly compressed beryllium.
Having unraveled electronic correlations in warm dense quantum plasmas, we open the path to study lights elements and potentially their mixtures 
for the extreme conditions encountered during inertial confinement fusion implosions and within astrophysical objects.
This will be a true game changer for a field that previously lacked predictive capability.

\bibliography{bibliography}
\bibliographystyle{Science}

\section*{Acknowledgments}

We gratefully acknowledge helpful feedback by Mandy Bethkenhagen.

This work was partially supported by the Center for Advanced Systems Understanding (CASUS) which is financed by Germany’s Federal Ministry of Education and Research (BMBF) and by the Saxon state government out of the State budget approved by the Saxon State Parliament.
This work has received funding from the European Research Council (ERC) under the European Union’s Horizon 2022 research and innovation programme
(Grant agreement No. 101076233, "PREXTREME"). 
Views and opinions expressed are however those of the authors only and do not necessarily reflect those of the European Union or the European Research Council Executive Agency. Neither the European Union nor the granting authority can be held responsible for them.
The work of Ti.~D., M.J.M., and F.R.G.~was performed under the auspices of the U.S. Department of Energy by Lawrence Livermore National Laboratory under Contract No. DE-AC52-07NA27344.
The work of T.G.~was partially supported by the European Union's Just Transition Fund (JTF) within the project \emph{R\"ontgenlaser-Optimierung der Laserfusion} (ROLF), contract number 5086999001, co-financed by the Saxon state government out of the State budget approved by the Saxon State Parliament.
The PIMC calculations were carried out at the Norddeutscher Verbund f\"ur Hoch- und H\"ochstleistungsrechnen (HLRN) under grant mvp00024, on a Bull Cluster at the Center for Information Services and High Performance Computing (ZIH) at Technische Universit\"at Dresden, and on the HoreKa supercomputer funded by the Ministry of Science, Research and the Arts Baden-W\"urttemberg and
by the Federal Ministry of Education and Research.

\newpage
\section*{Supplementary material}

\paragraph*{Model-free analysis of XRTS measurements.}

To extract the temperature $T$ and normalization $S_{ee}(q)=F_{ee}(q,0)$ from a measured XRTS signal without the need for models and approximations, we
switch to the imaginary-time domain~\cite{Dornheim_T_2022,Dornheim_insight_2022}, where the ITCF $F_{ee}(q,\tau)$ is connected to the dynamic structure factor $S_{ee}(q,\omega)$ via a two-sided Laplace transform,
\begin{eqnarray}\label{eq:Laplace}
    F_{ee}(q,\tau) = \mathcal{L}\left[S_{ee}(q,\omega)\right] = \int_{-\infty}^\infty \textnormal{d}\omega\ S_{ee}(q,\omega) e^{-\hbar\omega\tau}\ .
\end{eqnarray}
Working in the Laplace domain allows one to separate the physical information of interest from the properties of the combined source-and-instrument function $R(\omega)$ via the well-known convolution theorem~\cite{Dornheim_T2_2022,Dornheim_review},
\begin{eqnarray}\label{eq:convolution}
    \mathcal{L}\left[S_{ee}(q,\omega)\right] = \frac{\mathcal{L}\left[S_{ee}(q,\omega)\circledast R(\omega)\right]}{\mathcal{L}\left[R(\omega)\right]}\quad .
\end{eqnarray}
Given accurate knowledge of $R(\omega)$ based on either source monitoring or additional characterization studies~\cite{MacDonald_POP_2022}, it is straightforward to evaluate both the numerator and the denominator of Eq.~(\ref{eq:convolution}) even in the presence of experimental noise~\cite{Dornheim_T2_2022}.

A particularly useful application of Eqs.~(\ref{eq:Laplace}) and (\ref{eq:convolution}) is the symmetry relation of the ITCF, $F_{ee}(q,\beta-\tau)=F_{ee}(q,\tau)$, where $\beta=1/k_\textnormal{B}T$ is the inverse temperature. It is easy to see that the ITCF is symmetric around $\tau=\beta/2$, where it attains a minimum. This symmetry property is equivalent to the detailed balance relation of the dynamic structure factor $S_{ee}(q,-\omega)=e^{-\beta\hbar\omega}S_{ee}(q,\omega)$ that universally holds in thermodynamic equilibrium~\cite{quantum_theory}. This makes it directly possible to infer the temperature from the XRTS signal evaluated in the Laplace domain without any model calculations or approximations.

An additional obstacle is given by the necessarily finite detector range, whereas Eq.~(\ref{eq:Laplace}) in principle requires the integration over an infinite $\omega$-range. In practice, we define the two-sided symmetric incomplete Laplace transform
\begin{eqnarray}\label{eq:Laplace_truncated}
    \mathcal{L}_x\left[S_{ee}(q,\omega)\right] = \int_{-x}^x \textnormal{d}\omega\ S_{ee}(q,\omega) e^{-\hbar\omega\tau}\quad ,
\end{eqnarray}
whose convergence with $x$ is straightforward to check.

\begin{figure}\centering
\includegraphics[width=0.475\textwidth]{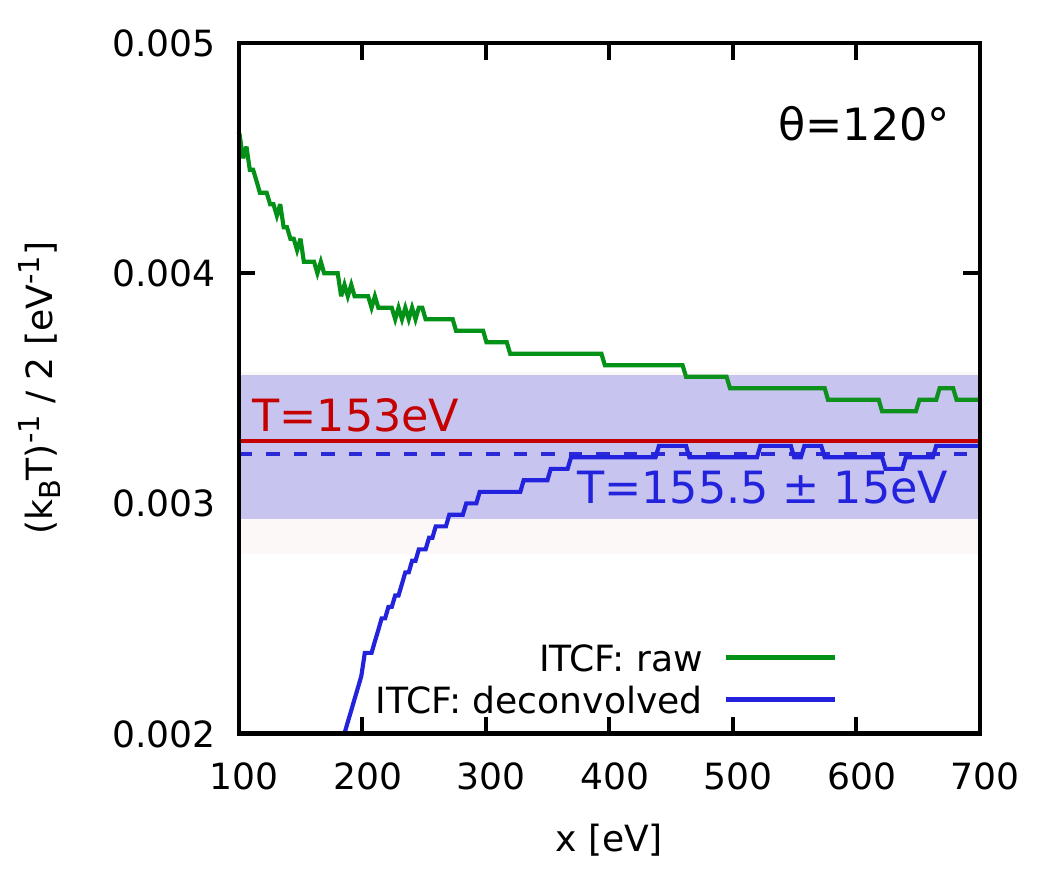}\includegraphics[width=0.475\textwidth]{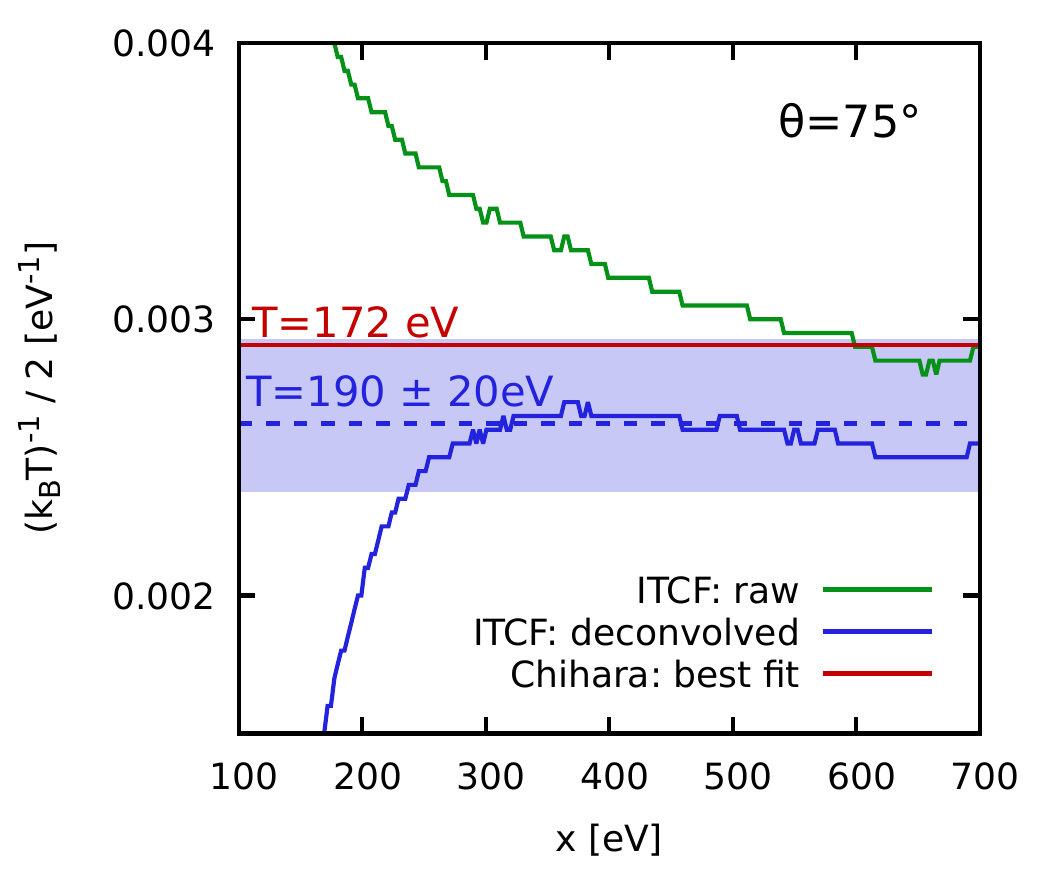}
\caption{\label{fig:ITCF_temperature}
Convergence of the temperature extracted from the ITCF-based analysis of NIF shots N170214(S3) and N170215(S3) with respect to the integration range $x$, see Eq.~(\ref{eq:Laplace_truncated}). (Green) blue: (not) including the source-and-instrument function $R(\omega)$; red: best fit from improved Chihara models~\cite{boehme2023evidence}; shaded blue area: uncertainty of the ITCF analysis.
}
\end{figure} 

In Fig.~\ref{fig:ITCF_temperature}, we show the corresponding ITCF-based temperature analysis of the two XRTS measurements at the NIF. For $\theta=120^{\circ}$, we find that the properly deconvolved data (blue) converge around a temperature of $T=155.5\,$eV, in excellent agreement with the improved Chihara model from Ref.~\cite{boehme2023evidence}. For $\theta=75^{\circ}$, the ITCF analysis gives us a temperature of $T=190\,$eV, with most of the associated uncertainty (shaded blue area) stemming from the influence of the source-and-instrument function.

\begin{figure}\centering
\includegraphics[width=0.475\textwidth]{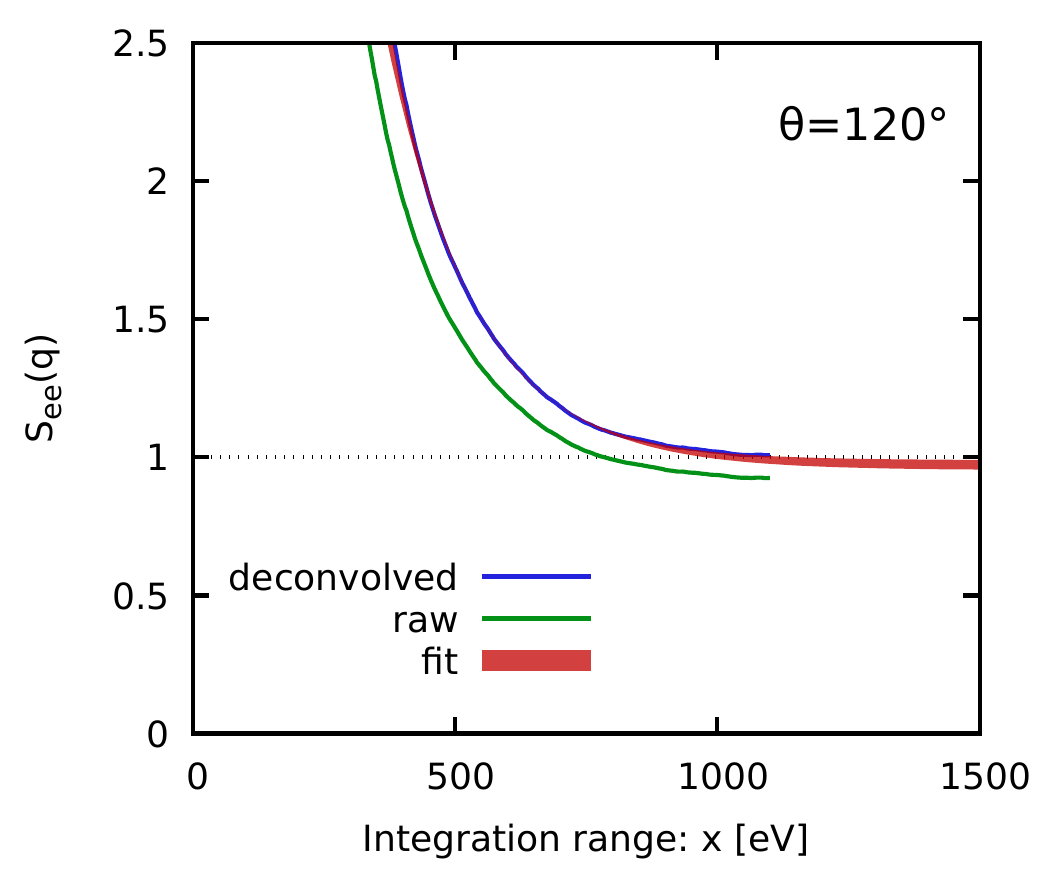}\includegraphics[width=0.475\textwidth]{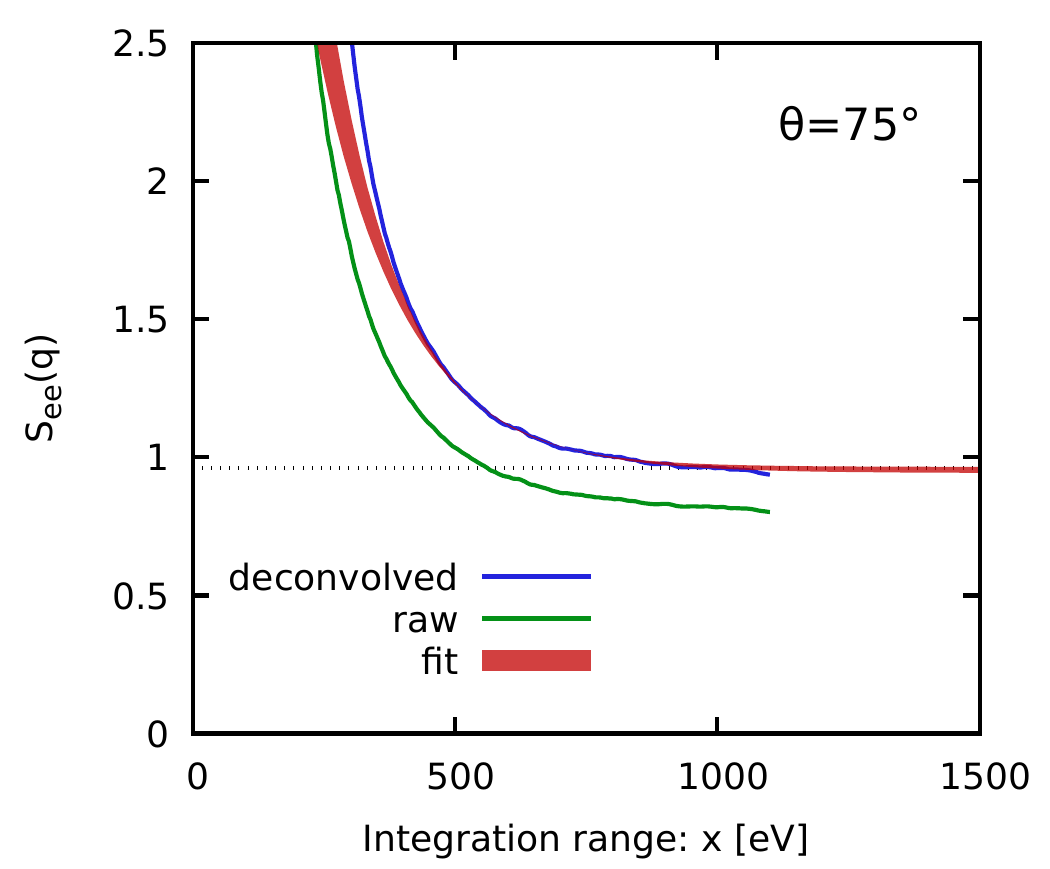}
\caption{\label{fig:ITCF_SSF}
Determination of the static structure factor $S_{ee}(q)$ (i.e. the normalization of the XRTS signal) from the imaginary-time f-sum rule as explained in Ref.~\cite{dornheim2023xray}.
}
\end{figure} 

Finally, Fig.~\ref{fig:ITCF_SSF} shows the extraction of the electronic SSF $S_{\mathrm{ee}}(q)$, i.e., the proper normalization of the XRTS signal, via the imaginary-time f-sum rule approach, recently introduced in Ref.~\cite{dornheim2023xray}. More specifically, we show the convergence of the extracted $S_{\mathrm{ee}}(q)$ with the symmetrically truncated integration range $x$ for the properly deconvolved (blue) and raw data (green). Evidently, the convergence behavior is well reproduced by the simple exponential ansatz
\begin{eqnarray}\label{eq:fit}
    f(x) = A + B e^{-Cx}\quad ,
\end{eqnarray}
which is shown as the dotted red curve. It is noted that Eq.~(\ref{eq:fit}) is well justified from a theoretical perspective by the expected exponential decay of the dynamic structure factor for large frequencies. In practice, the final estimate for $S_{\mathrm{ee}}(q)$ and the corresponding uncertainty (shaded red area in Fig.~\ref{fig:ITCF_SSF}) is obtained by performing several exponential fits over different reasonable intervals $x\in[x_\textnormal{min},x_\textnormal{max}]$ and by estimating the spread in the fitting parameter $A$.

\paragraph*{Ratio of elastic and inelastic contributions}

It is common practice to express the full electronic DSF as a sum of elastic and inelastic contributions~\cite{Vorberger_PRE_2015},
\begin{eqnarray}\label{eq:DSF_as_sum}
    S_{ee}(q,\omega) = \underbrace{W_R(q)\delta(\omega)}_{S_\textnormal{el}(q,\omega)} + S_\textnormal{inel}(q,\omega)\ .
\end{eqnarray}
We note that Eq.~(\ref{eq:DSF_as_sum}) does not presuppose any artificial decomposition into effectively bound and free electrons. Instead, the quasi-elastic contribution is due to the longer time scales of the ions due to their heavier mass; it is shaped as the source function $R(\omega)$ in the measured XRTS intensity.
From a theoretical perspective, it is straightforward to express the integrated ratio of $S_{el}(q,\omega)$ and $S_\textnormal{inel}(q,\omega)$ as
\begin{eqnarray}\label{eq:ratio}
  \frac{I_\textnormal{el}(q)}{I_\textnormal{inel}(q)} =  \frac{\int_{-\infty}^\infty \textnormal{d}\omega\ S_\textnormal{el}(q,\omega)}{ \int_{-\infty}^\infty \textnormal{d}\omega\ S_\textnormal{inel}(q,\omega) } = \frac{W_R(q)}{S_{ee}(q)-W_R(q)} = \left( \frac{S_{ee}(q)S_{II}(q)}{S_{eI}^2(q)}-1 \right)^{-1}\ .
\end{eqnarray}
The theoretical estimation of $I_\textnormal{el}(q)/I_\textnormal{inel}(q)$ thus requires explicit simulation results for correlation functions between all types of particles in the system.

\paragraph*{PIMC simulation details.}

We consider a fully spin-polarized system where $N_\uparrow = N_\downarrow = N/2$ with $N$ being the total number of electrons. Moreover, we consider effectively charge neutral systems where $N=Z_\textnormal{tot}N_\textnormal{Be}$ with $Z_\textnormal{tot}=4$ being the atomic charge and $N_\textnormal{Be}$ the total number of atoms.
The corresponding Hamiltonian governing the behaviour of the thus defined two-component plasma then reads
\begin{eqnarray}\label{eq:H}
    \hat{H} = &-& \frac{\hbar^2}{2m_e} \sum_{l=1}^N \nabla_l^2  - \frac{\hbar^2}{2m_I} \sum_{l=1}^{N_\textnormal{Be}} \nabla_l^2\\\nonumber &+& e^2 \left\{ \sum_{l<k}^N W_\textnormal{E}(\hat{r}_l,\hat{r}_k) + Z_\textnormal{tot}^2 \sum_{l<k}^{N_\textnormal{Be}} W_\textnormal{E}(\hat{I}_l,\hat{I}_k) - Z_\textnormal{tot} \sum_{k=1}^N\sum_{l=1}^{N_\textnormal{Be}} W_\textnormal{E}(\hat{I}_l,\hat{r}_k)\right\}\ .
\end{eqnarray}
The pair potential is given by the usual Ewald sum, where we follow the definitions of Fraser \emph{et al.}~\cite{Fraser_PRB_1996}.

The basic idea of the PIMC method~\cite{cep} is to express the canonical partition function (i.e., particle number $N$, volume $V$, and inverse temperature $\beta$ are fixed) in coordinate space,
\begin{eqnarray}\label{eq:Z}
Z_{N,V,\beta} = \frac{1}{N^\uparrow!N^\downarrow!} \sum_{\sigma_{N^\uparrow}\in S_{N^\uparrow}}\sum_{\sigma_{N^\downarrow}\in S_{N^\downarrow}} \xi^{N_\textnormal{pp}} \int_V \textnormal{d}\mathbf{R}\ \bra{\mathbf{R}} e^{-\beta\hat{H}} \ket{\hat{\pi}_{\sigma_{N^\uparrow}}\hat{\pi}_{\sigma_{N^\downarrow}}\mathbf{R}}\ .
\end{eqnarray}
The two sums over all possible permutations of the respective permutation groups (with $\hat{\pi}_{\sigma_{N^\uparrow}}$ and $\hat{\pi}_{\sigma_{N^\downarrow}}$ realizing a particular permutation) take into account that identical quantum particles cannot be distinguished. We further note that the vector $\mathbf{R}$ contains the coordinates of all electrons and ions in the system. In addition, $N_\textnormal{pp}$ is the number of pair exchanges required to realize a particular permutation, and $\xi=1$, $\xi=0$, and $\xi=-1$ correspond to the physically meaningful cases of Bose-Einstein, Maxwell-Boltzmann, and Fermi-Dirac statistics, with the latter governing the behavior of the electrons in WDM. For completeness, we note that any exchange effects of the ions can safely be neglected at the conditions that are of interest in the present work.

\begin{figure}\centering
\includegraphics[width=0.475\textwidth]{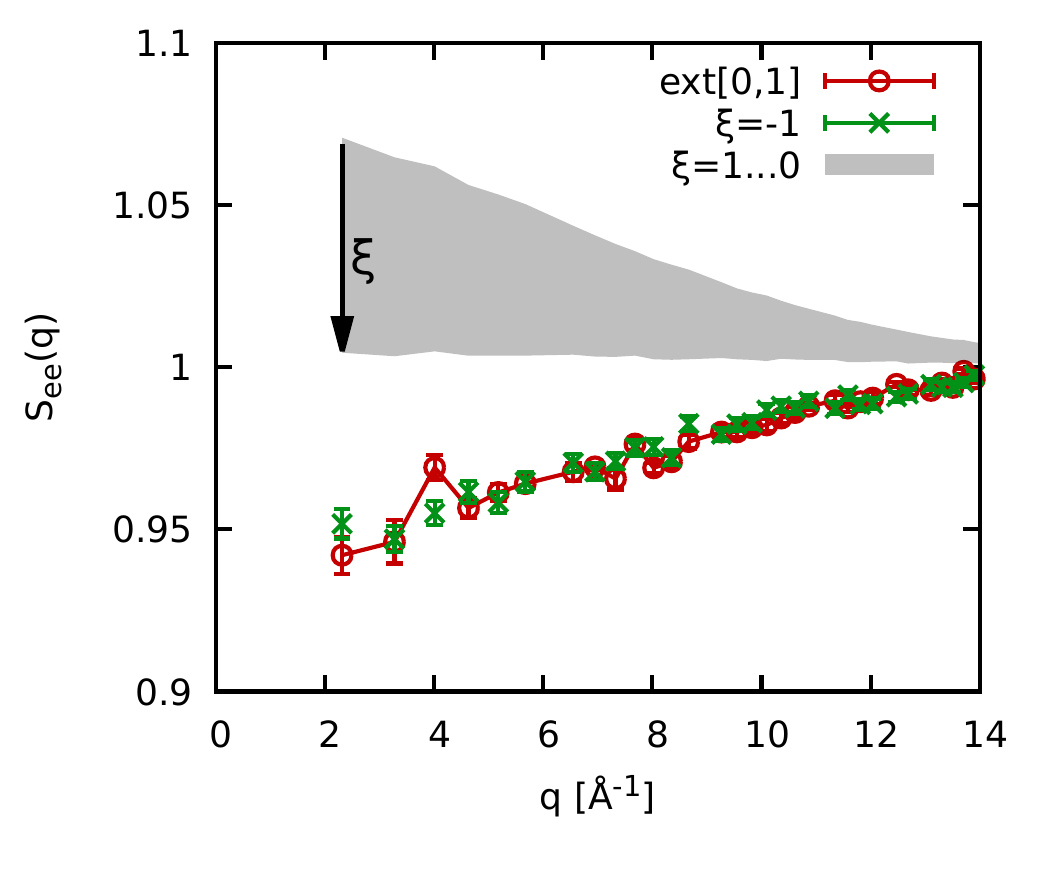}
\includegraphics[width=0.475\textwidth]{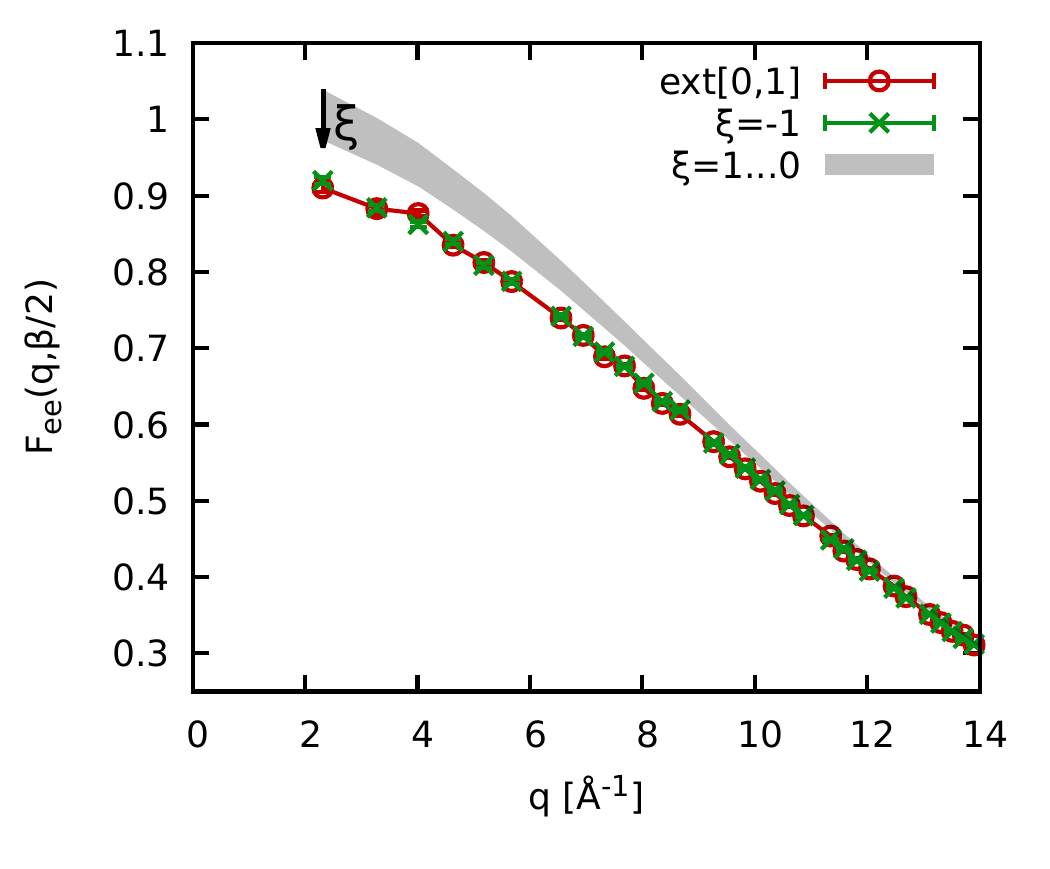}\\
\includegraphics[width=0.475\textwidth]{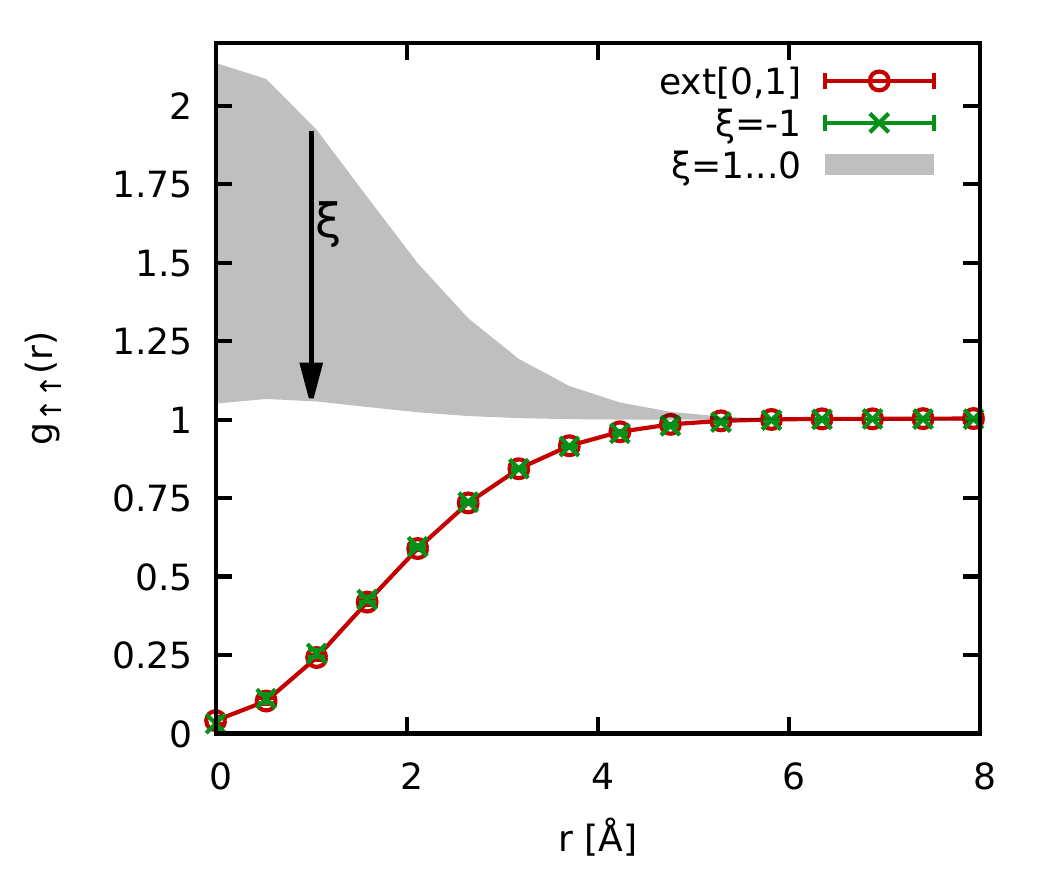}\includegraphics[width=0.475\textwidth]{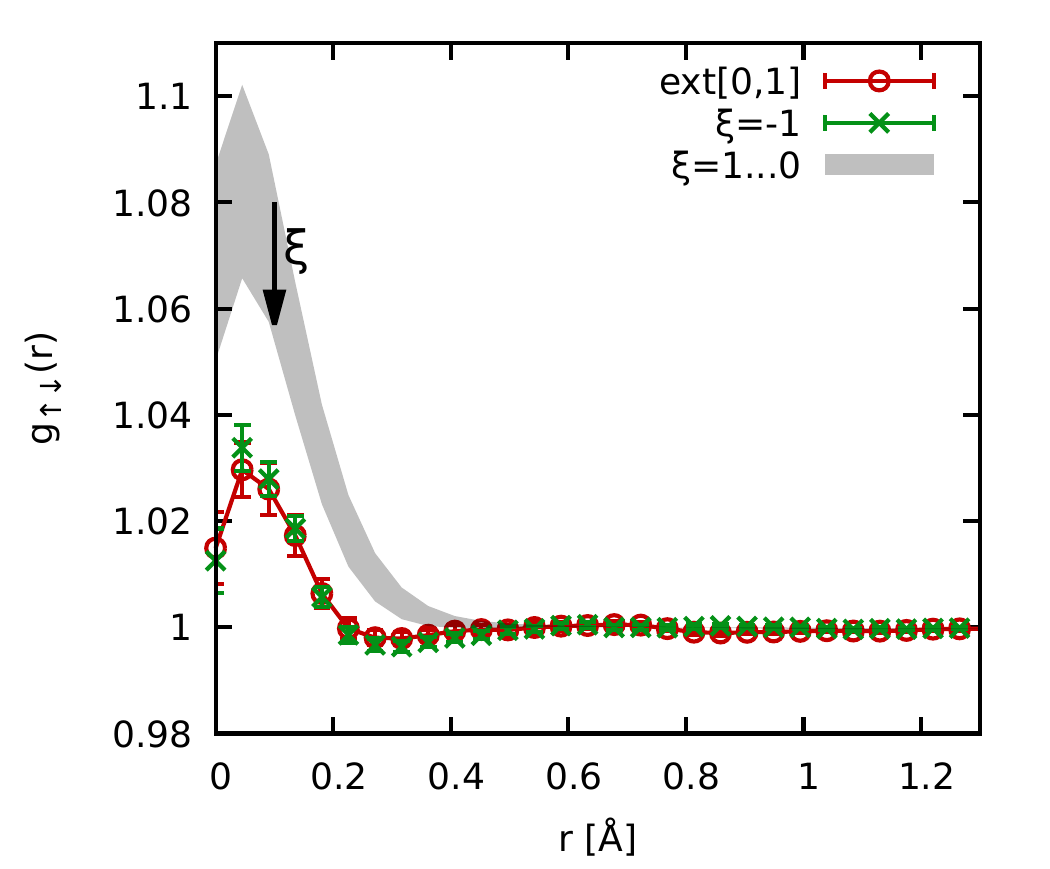}
\caption{\label{fig:FinalExtrapolationN170214}  
Benchmarking the $\xi$-extrapolation approach for $N_\textnormal{Be}=10$ Be atoms at $T=155.5\,$eV and $r_s=0.93$ ($\rho=7.5\,$g/cc). Green crosses: exact fermionic PIMC results (i.e., $\xi=-1$); shaded grey area: PIMC results in the sign-problem free domain of $\xi\in[0,1]$; red circles: quadratic extrapolation to the fermionic limit.
}
\end{figure} 

For $\xi=-1$, it is well known that positive and negative contributions to any observable cancel to a large degree, which leads to an exponential increase in the required compute time with increasing the number of electrons $N=N^\uparrow+N^\downarrow$ or decreasing the temperature $T$; this is the notorious fermion sign problem~\cite{troyer,dornheim_sign_problem}. Here, we follow the approach introduced in Refs.~\cite{Xiong_JCP_2022,dornheim2023fermionic} and consider the general case of a continuous variable $\xi\in[-1,1]$. The basic idea is to carry out PIMC simulations in the sign-problem free domain of $\xi\geq0$, and to quadratically extrapolate to the fermionic limit of $\xi=-1$. Indeed, Dornheim \emph{et al.}~\cite{dornheim2023fermionic,Dornheim_JPCL_2024} have shown very recently that this allows for quasi-exact uniform electron gas results for a range of observables including the SSF and ITCF over substantial parts of the WDM regime.  A particular strength of this methodology is that it allows one to rigorously assess its accuracy for the case of a small system for which simulations can be performed even in the fermionic limit of $\xi=-1$. Its continued reliability for substantially larger number of particles is then guaranteed both empirically~\cite{dornheim2023fermionic}, and from the principle of electronic nearsightedness~\cite{Kohn_PNAS_2005}. In essence, this approach allows for unprecedented PIMC simulations of fermions without the exponential computational bottleneck, without any uncontrolled approximations or empirical parameters, and without the restrictions on the sampling of the imaginary-time structure inherent to the fixed-node approximation~\cite{Ceperley1991}.

In Fig.~\ref{fig:FinalExtrapolationN170214}, we show this extrapolation for $N_\textnormal{Be}=10$ Be atoms (i.e., $N=40$ electrons) at $r_s=0.93$ ($\rho=7.5\,$g/cc) and $T=155.5\,$eV. The top left panel shows the $q$-dependence of the electronic SSF $S_{\mathrm{ee}}(q)$. More specifically, the green crosses show direct PIMC results for $\xi=-1$ that are subject to the full sign problem. We find an average sign of $S\approx0.098$, which means that the simulations are computationally involved, but still feasible. 
In addition, the shaded grey area encompasses our sign-problem free PIMC results for the range $1\geq\xi\geq0$. Evidently, the bosonic SSF exhibits the opposite trend compared to the fermions with respect to $q$. Nevertheless, the quadratic extrapolation that is based on the grey area accurately reproduces the fermionic limit, see the red circles. The same trend is discerned for the thermal structure factor $S^{\beta/2}_{\mathrm{ee}}(q)=F_{\mathrm{ee}}(q,\beta/2)$, as shown in the top right panel of Fig.~\ref{fig:FinalExtrapolationN170214}, although the effects of quantum statistics are less pronounced for $\tau=\beta/2$.

In the bottom row, we repeat this analysis for the spin-resolved electronic PCF, with the left and right panels showing results for the spin-diagonal and spin-offdiagonal component. For $g_{\uparrow\uparrow}(r)$, spin effects predominantly shape the behavior for $r\lesssim4\,$\AA; fermions exhibit the familiar exchange--correlation hole with $g_{\uparrow\uparrow}(0)=0$, whereas bosons tend to cluster around each other exhibiting the opposite trend. Nevertheless, the extrapolation works exceptionally well and reproduces the fermionic curve. For $g_{\uparrow\downarrow}(r)$, we find a more subtle behavior and spin-effects play a less important role. At the same time, the $\xi$-extrapolation method cannot be distinguished from the fermionic benchmark results within the (small) Monte Carlo error bars.

\paragraph*{Convergence with number of imaginary-time propagators.} An indispensable ingredient to the PIMC method is given by the factorization of the density operator $e^{-\beta\hat{H}}\neq e^{-\beta\hat{W}}e^{-\beta\hat{K}}$, where $\hat{W}$ and $\hat{K}$ are the potential and kinetic energy operators. Here, we use an implementation of the pair approximation, as described in Ref.~\cite{Boehme_PRE_2023}, that becomes exact in the limit of $P\to\infty$ as $\sim1/P^4$, where $P$ is the number of imaginary-time propagators, or, equivalently, the number of high-temperature factors.

\begin{figure}\centering
\includegraphics[width=0.475\textwidth]{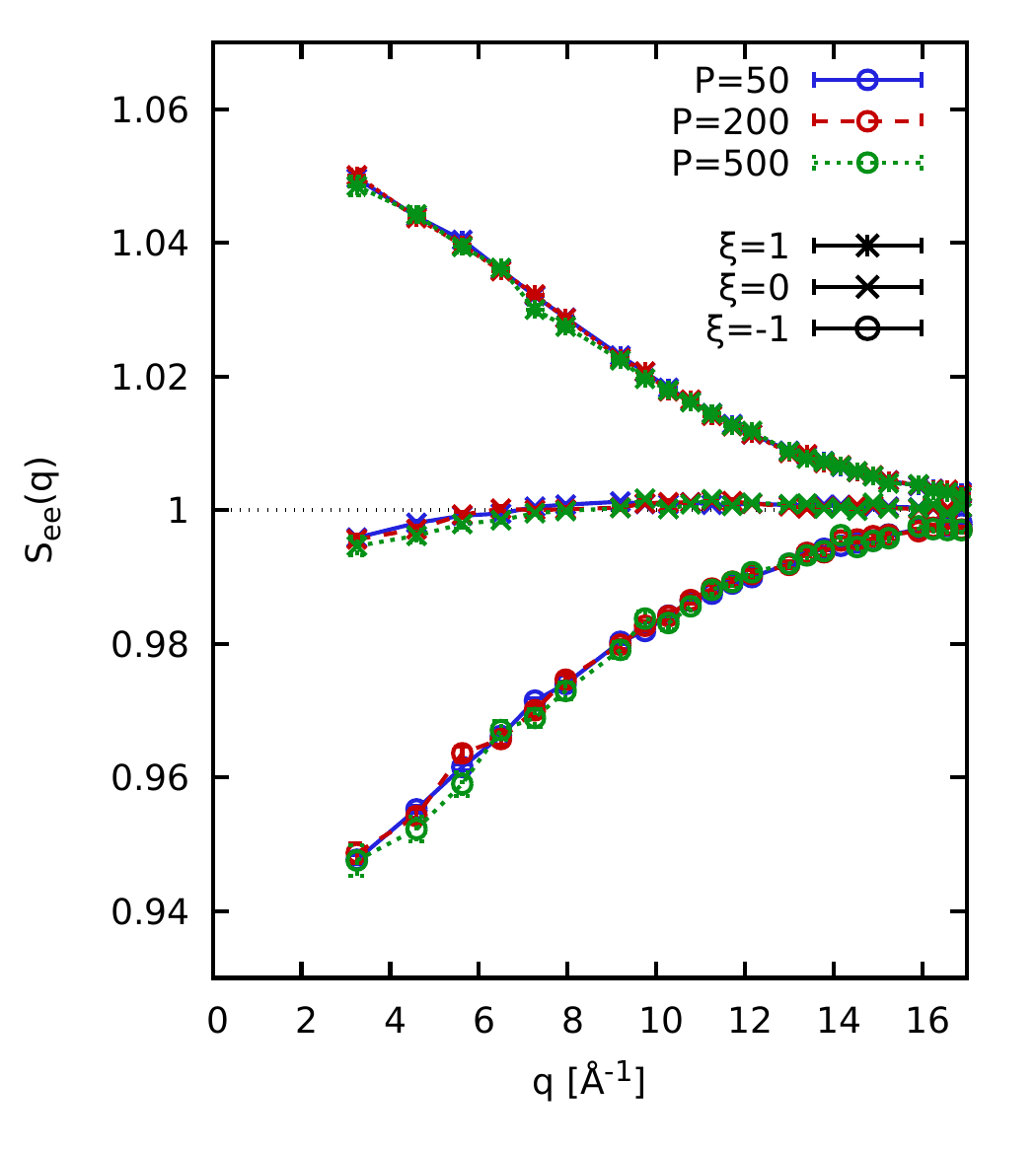}
\includegraphics[width=0.475\textwidth]{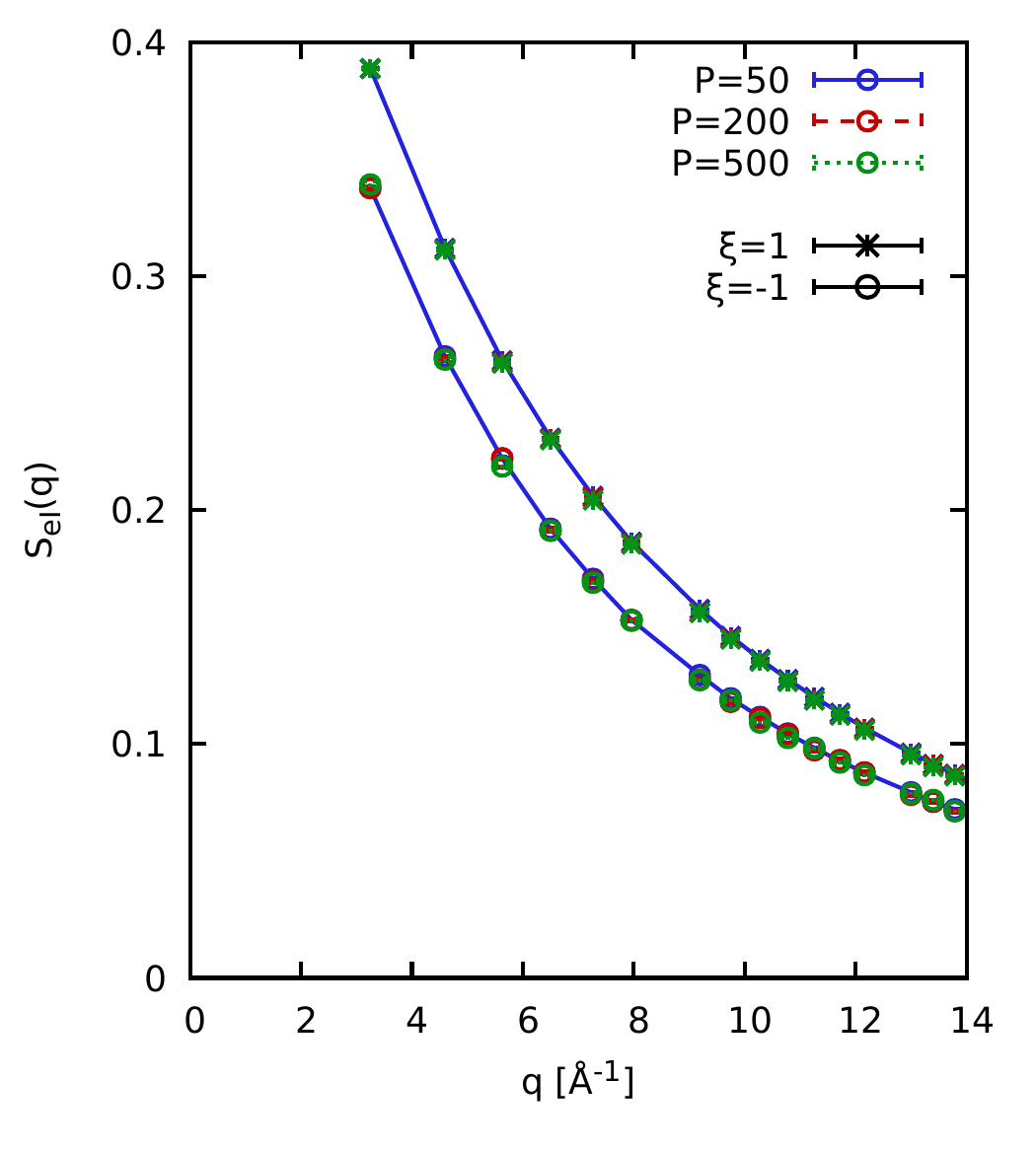}
\caption{\label{fig:P_convergence_SSF}
Convergence of $S_{ee}(q)$ and $S_{eI}(q)$ with the number of imaginary-time propagators for $N_\textnormal{Be}=4$, $\rho=7.58\,$g/cc, and $T=155.5\,$eV.
}
\end{figure} 

In Fig.~\ref{fig:P_convergence_SSF}, we show results for the electronic SSF $S_{\mathrm{ee}}(q)$ [left] and the electron-ion SSF $S_{\mathrm{eI}}(q)$ [right] for $N_\textnormal{Be}=4$ Be atoms at $r_s=0.9$ ($\rho=7.58\,$g/cc) and $T=155.5\,$eV. In particular, the stars, crosses, and circles correspond to the relevant cases of $\xi=1$, $\xi=0$, and $\xi=-1$, respectively, and the blue, red, and green color distinguishes simulation results for $P=50$, $P=200$, and $P=500$. Evidently, we find no systematic dependence even for $P=50$. In practice, we use $P=200$ throughout this work, as this has the beneficial side effect of a good $\tau$-resolution for the ITCF $F_{ee}(q,\tau)$.

\end{document}